# THE CONTRIBUTION OF SOVIET SCIENTISTS TO THE EMERGENCE OF HIGH-ENERGY NEUTRINO ASTROPHYSICS: FROM THE CHERENKOV EFFECT TO UNDERWATER AND IN-ICE TELESCOPES

**Rizoi Bakhromzod[1,2]***

[1]S. U. Umarov Physical-Technical Institute, National Academy of Sciences of Tajikistan, 299/1 Aini Street, Dushanbe 734063, Republic of Tajikistan

[2]Institute of Astrophysics, National Academy of Sciences of Tajikistan 299/5 Aini Street, Dushanbe 734063, Republic of Tajikistan

*e-mail: rizo@physics.msu.ru, bakhromzod@gmail.com

## Abstract

The 2026 Nobel Prize in Physics was awarded to Francis Halzen “for decisive contributions to the IceCube Neutrino Observatory and the discovery of high-energy neutrinos of astrophysical origin.” This article examines the Soviet contribution to the broader historical development of high-energy neutrino astrophysics. Four interconnected strands are distinguished: the physics of Cherenkov radiation and its application to cosmic rays; early proposals for large-scale neutrino detection underground and in natural water; the theory of atmospheric, cosmogenic, and astrophysical neutrinos; and the development of underground, underwater, radio, and in-ice detection techniques. The article shows that M. A. Markov’s 1960 proposal was concerned primarily with atmospheric neutrinos and with detecting charged secondary particles through Cherenkov radiation in water, whereas the use of Antarctic ice was initially developed in the Soviet Union in the radio domain. The optical concept of a detector in deep polar ice was formulated by F. Halzen and J. G. Learned in 1988. The Pamir cosmic-ray school, the theoretical work of G. T. Zatsepin, V. A. Kuzmin, and V. S. Berezinsky, the Baksan Neutrino Observatory, Soviet participation in DUMAND, and the formation of the Lake Baikal neutrino program are considered separately. The Pamir material is included as historical and methodological context; it does not imply a direct experimental lineage from Pamir installations to IceCube. The historical links of these strands to IceCube are therefore different in character: conceptual and citation-based in some cases, methodological or institutional in others. This distinction helps separate documented continuity of ideas from the broader scientific context in which modern high-energy neutrino astronomy emerged.

## Introduction

On 6 October 2026, the Royal Swedish Academy of Sciences announced the award of the Nobel Prize in Physics to Francis Halzen for decisive contributions to the IceCube Neutrino Observatory and the discovery of high-energy neutrinos of astrophysical origin (Nobel Committee for Physics, 2026; Royal Swedish Academy of Sciences, 2026). The award draws attention not only to the history of IceCube itself but also to the much longer development of ideas that made high-energy neutrino astronomy possible.

The prehistory of IceCube cannot be reduced to a single uninterrupted chain from one early experiment to the present observatory. Several lines converged: the study of the Cherenkov effect, cosmic-ray physics, calculations of atmospheric and cosmogenic neutrinos, the development of underground detectors, the design of deep-water telescopes, and the search for viable detection methods in polar ice. Soviet scientists contributed to each of these lines, but the nature of their connection to IceCube differs. In some cases the relationship is one of direct conceptual precedence; in others it is a parallel experimental tradition or the formulation of physical questions that later became central to neutrino astrophysics.

The aim of this article is to reconstruct these lines without retrospectively attributing to early work ideas that were not present in the original sources. Particular attention is therefore paid to the distinction between atmospheric and astrophysical neutrinos, between underground and underwater detection, and between Soviet radio concepts for Antarctic ice and the later optical AMANDA–IceCube line.

## 1. IceCube and the discovery of astrophysical neutrinos

IceCube is a cubic-kilometre optical neutrino detector that uses transparent Antarctic ice as the detection medium. Its photomultiplier modules register Cherenkov light from charged particles produced in neutrino interactions. Construction of the completed configuration was finished in 2011, and the detector and its online systems are described in detail by the IceCube Collaboration (Aartsen et al., 2017).

In 2013, the collaboration first reported two events with energies of about 1 PeV (Aartsen et al., 2013a) and then a statistically significant excess of high-energy events incompatible with a purely atmospheric origin (Aartsen et al., 2013b). This result established the existence of an extraterrestrial high-energy neutrino flux. Neutrino astronomy subsequently progressed from the discovery of a diffuse flux toward the identification of sources. Important milestones included observations associated with the blazar TXS 0506+056 (Aartsen et al., 2018a, 2018b), evidence for neutrino emission from the direction of NGC 1068 (IceCube Collaboration, 2022), and the observation of high-energy neutrinos from the Galactic plane (IceCube Collaboration, 2023).

The official 2026 Nobel materials place Halzen’s work within the longer history of large-scale neutrino telescopes, including DUMAND and underwater projects (Nobel Committee for Physics, 2026). For assessing the Soviet contribution, it is essential to distinguish the early use of natural water as a Cherenkov detection medium from the later proposal to use deep polar ice for an optical detector.

## 2. Cherenkov radiation and the cosmic-ray school

### 2.1. Physical basis of the method

Optical neutrino telescopes detect Cherenkov radiation from secondary charged particles. The experimental discovery of the visible radiation produced by fast charged particles in transparent media is associated with the work of P. A. Cherenkov, while the theoretical description was developed by I. E. Tamm and I. M. Frank (Cherenkov, 1934; Tamm and Frank, 1937). In 1958, Cherenkov, Frank, and Tamm received the Nobel Prize in Physics for the discovery and interpretation of this effect (NobelPrize.org, 1958).

The next important step for high-energy astrophysics was the use of Cherenkov light in cosmic-ray studies. In 1953, W. Galbraith and J. V. Jelley detected short light pulses associated with extensive air showers (Galbraith and Jelley, 1953). In the Soviet Union, A. E. Chudakov and N. M. Nesterova developed this approach at the high-altitude Pamir station of the Lebedev Physical Institute. Their mid-1950s experiments combined optical receivers with parabolic mirrors and photomultipliers with a hodoscope of Geiger counters; the main campaign was carried out at an altitude of 3860 m (Nesterova and Chudakov, 1955; Chudakov and Nesterova, 1958).

### 2.2. Pamir: significance and limits of its connection to neutrino astronomy

Pamir research occupies an important place in the history of Soviet cosmic-ray physics. Lebedev Institute expeditions to the Pamirs began in the 1940s, and a major review of extensive air-shower results was published by N. A. Dobrotin, G. T. Zatsepin, and colleagues in 1953 (Skobeltsyn, 1946; Dobrotin et al., 1953). These studies helped form an experimental school combining high-altitude measurements, secondary-particle detection, and the study of hadronic interactions.

The later Pamir experiment, begun in 1971, was a different program. It used large X-ray emulsion chambers to investigate cosmic-ray interactions at energies of 10^14–10^17 eV. Participating institutions included the Lebedev Institute, the Institute for Nuclear Research of the USSR Academy of Sciences, Moscow State University, institutes in several Soviet republics and Poland, including the S. U. Umarov Physical-Technical Institute of the Academy of Sciences of the Tajik SSR (Baiburina et al., 1984). This program is important for the history of cosmic-ray physics, but it was not a direct experimental predecessor of IceCube.

The Pamir connection to neutrino astronomy should therefore be formulated as methodological and personal rather than as a direct lineage of detector construction. Chudakov later took part in DUMAND discussions and, in 1980, proposed Lake Baikal as a site for a Soviet deep-water neutrino program (Spiering, 2012, p. 528). Zatsepin's scientific trajectory likewise linked cosmic-ray physics to neutrino physics: he was a co-author of the 1953 air-shower review and later of foundational calculations of atmospheric neutrino production and the ultra-high-energy cosmic-ray cutoff (Dobrotin et al., 1953; Zatsepin and Kuzmin, 1961, 1966). The Pamir installations themselves, however, did not detect astrophysical neutrinos and did not evolve directly into the IceCube detector.

## 3. M. A. Markov and the idea of large-scale neutrino detection

### 3.1. The 1960 Rochester conference

A key early document is M. A. Markov's paper "On high energy neutrino physics," presented at the 1960 Rochester conference. In the section on cosmic-ray experiments, Markov proposed placing apparatus "in an underground lake or deep in the ocean" in order to determine the directions of charged particles by Cherenkov radiation (Markov, 1960). This statement contains the essential idea of using a large volume of a natural transparent medium as part of a neutrino detector.

The historical context of the proposal must be kept precise. The main subject of Markov's discussion was neutrinos produced by meson decays in the Earth's atmosphere. Extra-atmospheric, including Galactic, neutrinos were mentioned as a possible extension of the program, but they were not the primary experimental target of the 1960 proposal. It is therefore appropriate to describe the Rochester idea as an early conceptual predecessor of underwater optical neutrino telescopes, but not as a ready-made project for an astrophysical observatory in ice.

In 1961, Markov and I. M. Zheleznykh discussed underground detection of high-energy cosmic-ray neutrinos, particularly through upward-going muons (Markov and Zheleznykh, 1961). This paper became one of the early reference points for later treatments of high-energy neutrino detection; it was explicitly cited in the review by Gaisser, Halzen, and Stanev (1995).

### 3.2. From water to ocean, atmosphere, and ice

By the mid-1980s, Markov and Zheleznykh were considering several media and detection techniques. Their 1986 paper "Large-scale Cherenkov detectors in ocean, atmosphere and ice" discussed deep-water optical detection, large-volume radio detection in ice, and other approaches to high-energy neutrino detection (Markov and Zheleznykh, 1986). The concept of a giant neutrino detector had thus broadened far beyond the original underground and underwater scheme.

The chronology nevertheless remains important. In 1960 Markov proposed water and the Cherenkov method; radio detection in Antarctic ice developed later; and the optical concept of a high-energy neutrino detector in deep polar ice was formulated by Halzen and Learned in 1988. Collapsing these stages into a single early proposal would create the misleading impression that the modern IceCube concept had already been formulated in Markov's first work.

## 4. The theoretical line: atmospheric and cosmogenic neutrinos

### 4.1. Atmospheric-neutrino calculations

The practical development of underground neutrino experiments required quantitative estimates of background and signal fluxes. In 1961, G. T. Zatsepin and V. A. Kuzmin published a calculation of neutrino production in the atmosphere, relating neutrino spectra to cascade processes and the decays of secondary mesons (Zatsepin and Kuzmin, 1961). Together with the work of Markov and Zheleznykh from the same period, these calculations helped establish the theoretical basis for the analysis of atmospheric neutrinos.

Atmospheric neutrinos later became both a research subject in their own right and the principal irreducible background for high-energy neutrino astronomy. In modern telescopes such as IceCube, accurate modeling of the atmospheric component is essential for isolating an astrophysical signal.

### 4.2. The GZK effect and cosmogenic neutrinos

Following the discovery of the cosmic microwave background, K. Greisen and independently G. T. Zatsepin and V. A. Kuzmin showed that ultra-high-energy cosmic-ray protons should lose energy in interactions with background photons (Greisen, 1966; Zatsepin and Kuzmin, 1966). This effect is known as the GZK cutoff.

In 1969, V. S. Berezinsky and G. T. Zatsepin connected these interactions with the production of ultra-high-energy neutrinos (Berezinsky and Zatsepin, 1969). V. S. Berezinsky and A. Yu. Smirnov later examined such neutrino fluxes and the prospects for their detection (Berezinsky and Smirnov, 1975), while related estimates were developed by other authors, including Stecker (1973). These studies provided one of the major physical motivations for constructing very-large-volume detectors.

By the 1970s, neutrino astrophysics was emerging in Soviet literature as a distinct research program. Berezinsky and Zatsepin discussed experiments with very-high-energy cosmic neutrinos and the DUMAND project (Berezinsky and Zatsepin, 1977), and Berezinsky later systematized the problem of high-energy neutrino sources and fluxes (Berezinsky, 1981). This theoretical line remains directly visible in modern cosmogenic-neutrino studies (Ahlers et al., 2010) and in Halzen's historical and scientific reviews (Halzen, 2014).

## 5. Underground experiments: Baksan

The underground branch of neutrino physics addressed a different but closely related problem: suppressing the intense atmospheric-muon background with overburden and identifying neutrino-induced particles. In the Soviet Union, the Baksan Neutrino Observatory of the Institute for Nuclear Research became a major center of this work. The Baksan Underground Scintillation Telescope began operation in 1977 and was used for a broad program in cosmic-ray and neutrino physics (Spiering, 2021).

Baksan's best-known astrophysical result was the detection of five neutrino events associated with SN 1987A within 9.1 s (Alexeyev et al., 1988). The burst was also observed by Kamiokande-II and IMB (Hirata et al., 1987; Bionta et al., 1987). These events had energies of order tens of MeV, many orders of magnitude below the energies characteristic of IceCube's principal astrophysical program. Baksan and IceCube therefore belong to the same broad history of neutrino astronomy but operate in different energy regimes and with different detector technologies.

For IceCube, the experience of underground experiments is methodologically relevant in such areas as background rejection, directional reconstruction, and statistical identification of neutrino events. There is, however, no direct constructional lineage from Baksan to IceCube.

## 6. DUMAND and Lake Baikal: the underwater line

### 6.1. DUMAND

DUMAND (Deep Underwater Muon and Neutrino Detector) was the first major international project for a deep-water high-energy neutrino telescope. It developed the principle anticipated by Markov: a large volume of transparent natural water acts both as target material and as the medium through which Cherenkov photons propagate. Soviet scientists involved in DUMAND discussions included A. E. Chudakov, V. S. Berezinsky, L. B. Bezrukov, B. A. Dolgoshein, I. M. Zheleznykh, and others (Spiering, 2012, p. 528).

According to Spiering's historical reconstruction, Soviet participation in DUMAND was effectively terminated in 1980 in the context of the political tensions that followed the Soviet intervention in Afghanistan (Spiering, 2012, p. 528). The DUMAND project itself continued until 1995 but was never completed as a full neutrino telescope. Its importance lies in the development of deep-water detection technology and in the formation of the international high-energy neutrino-astronomy community.

### 6.2. The Lake Baikal program

In parallel, the Soviet Union developed its own underwater program. Spiering records that in 1980 Chudakov proposed the deep water of Lake Baikal as the site for a "Soviet DUMAND" (Spiering, 2012, p. 528). The organizational development of the program was associated particularly with G. V. Domogatsky, L. B. Bezrukov, and the Institute for Nuclear Research team. A stationary string of optical modules was deployed in 1984, and the NT200 telescope was subsequently built during the 1990s.

Baikal NT200 became a major underwater neutrino installation and demonstrated the viability of the deep-water Cherenkov method. Its design and early results were reported by Belolaptikov et al. (1997), atmospheric neutrino detection with NT-96 by Balkanov et al. (1999), and the broader development of the Baikal detector program has been reviewed by Avrorin et al. (2022).

The modern continuation of this program is Baikal-GVD. Using data from 2018–2021, the collaboration reported a measurement of the diffuse astrophysical neutrino flux with an excess at approximately the 3$\sigma$ level (Allakhverdyan et al., 2023). An analysis of six seasons of data from 2018–2024, released as a preprint in 2025, reports a statistical significance of 5.1$\sigma$ (Allakhverdyan et al., 2025). The Lake Baikal line has therefore become an independent modern measurement of the diffuse astrophysical neutrino flux first established by IceCube.

## 7. Antarctic ice: from radio detection to an optical telescope

### 7.1. The radio line

G. A. Askaryan showed that an electromagnetic cascade in a dense medium develops an excess negative charge and can produce coherent radio emission (Askaryan, 1961). This result provided the physical basis for radio detection of ultra-high-energy particle cascades.

In the early 1980s, G. A. Gusev and I. M. Zheleznykh considered the detection of neutrinos and muons through radio emission from cascades in natural dielectric media, explicitly identifying Antarctic ice as an attractive medium because of the enormous radio-transparent target volume (Gusev and Zheleznykh, 1984). In 1986, Markov and Zheleznykh summarized prospects for large-scale detectors in ocean, atmosphere, and ice (Markov and Zheleznykh, 1986).

This line was technologically different from the future IceCube detector: the proposed signal was radio emission from particle cascades rather than optical Cherenkov light recorded by photomultipliers. Soviet radio work in ice should therefore be regarded as an important preceding research direction and as part of the context in which Antarctic ice became a detector medium, but not as a ready-made prototype of IceCube.

### 7.2. The optical detector in ice

The decisive step toward an optical in-ice telescope was the 1988 paper by F. Halzen and J. G. Learned, “High Energy Neutrino Detection in Deep Polar Ice” (Halzen and Learned, 1988). It proposed deep polar ice as a transparent medium for optical detection of Cherenkov light from neutrino-induced particles. Spiering’s historical review connects the origin of this idea to a 1988 discussion in which Halzen learned of plans for a Soviet radio test array at Vostok Station; the high threshold of the radio approach prompted consideration of the DUMAND optical principle in ice (Spiering, 2012, p. 537).

By the beginning of the 1990s, muon detection using polar ice as a Cherenkov medium had been experimentally demonstrated (Lowder et al., 1991). The AMANDA project followed, and its operating principle and early results were reported by Andres et al. (2000). IceCube was the large-scale continuation of this optical in-ice line.

The historical sequence can therefore be summarized as follows: Markov’s early proposal — a large detector in water; DUMAND and Lake Baikal — development of deep-water optical detection; Soviet radio work — exploration of Antarctic ice in the radio domain; Halzen and Learned — transition to optical detection in deep polar ice; AMANDA — experimental validation; IceCube — a cubic-kilometre neutrino observatory.

## 8. Comparative assessment of the main lines

For historical analysis it is useful to distinguish among the physical basis of the method, detector concepts, the formulation of astrophysical objectives, and institutional or experimental continuity. Table 1 summarizes the chronology, while Table 2 characterizes the relation of the principal Soviet lines to IceCube.

| Year | Event | Main participants / source |
|---|---|---|
| 1934–1937 | Discovery and theory of Cherenkov radiation | Cherenkov; Tamm and Frank |
| 1944–1950s | High-altitude cosmic-ray research in the Pamirs | Lebedev Institute; Skobeltsyn; Dobrotin; Zatsepin |
| 1953–1958 | Detection of Cherenkov light from extensive air showers | Galbraith and Jelley; Chudakov and Nesterova |
| 1960 | Proposal for a large detector in a lake/ocean using Cherenkov detection | M. A. Markov |
| 1961 | Underground detection of high-energy neutrinos; atmospheric-neutrino calculations | Markov and Zheleznykh; Zatsepin and Kuzmin |
| 1966–1969 | GZK cutoff and cosmogenic neutrinos | Greisen; Zatsepin and Kuzmin; Berezinsky and Zatsepin |
| 1970s | Development of neutrino astrophysics; DUMAND | Berezinsky and Zatsepin; international collaboration |
| 1977 | Baksan Underground Scintillation Telescope begins operation | Institute for Nuclear Research, USSR Academy of Sciences |
| 1980–1984 | Formation of the Lake Baikal program; first stationary optical systems | Chudakov; Domogatsky; Bezrukov et al. |
| 1984–1986 | Radio detection of neutrinos in natural dielectric media, including Antarctic ice | Gusev and Zheleznykh; Markov and Zheleznykh |
| 1987 | Detection of SN 1987A neutrinos | Baksan; Kamiokande-II; IMB |
| 1988 | Optical detection of high-energy neutrinos in deep polar ice proposed | Halzen and Learned |
| 1990s–2000s | AMANDA; development of Baikal NT200 | International collaborations |
| 2011–2013 | Completion of IceCube; discovery of diffuse astrophysical neutrino flux | IceCube Collaboration |
| 2023–2025 | Diffuse-flux measurements with Baikal-GVD | Baikal-GVD Collaboration |
| 2026 | Nobel Prize in Physics awarded to F. Halzen | Royal Swedish Academy of Sciences |

*Table 1. Chronology of the principal developments discussed in the article.*

| Line of research | Type of connection | Assessment |
|---|---|---|
| Cherenkov effect (Cherenkov, Tamm, Frank) | Physical basis | Direct: optical neutrino telescopes detect Cherenkov radiation from secondary charged particles. |
| Pamir Cherenkov experiments | Methodological | Early development of Cherenkov methods in cosmic-ray physics; no direct constructional lineage to IceCube. |
| Markov, 1960 | Conceptual | Early proposal for a large detector in natural water with directional reconstruction using Cherenkov light. |
| Zatsepin–Kuzmin; Berezinsky–Zatsepin | Theoretical | Atmospheric- and cosmogenic-neutrino calculations and formulation of major scientific objectives for large detectors. |

| Line of research | Type of connection | Assessment |
|---|---|---|
| Baksan | Thematic and methodological | Underground neutrino physics and SN 1987A neutrino astronomy; different energy range and detector technology. |
| DUMAND → Lake Baikal | Experimental and institutional | Development of deep-water optical detection; Baikal-GVD is now an independent operating observatory. |
| Soviet radio work in ice | Preceding technological context | Demonstrated the potential of polar ice for large-volume radio detection; method differs from optical IceCube. |
| Halzen–Learned → AMANDA → IceCube | Direct constructional | Optical in-ice line that led directly to IceCube. |

*Table 2. Types of historical connection between major Soviet research lines and IceCube.*

## 9. Discussion

The history of high-energy neutrino astronomy shows that major scientific instruments rarely arise from a single invention. IceCube brought together a physical effect discovered and explained in the 1930s, the concept of a detector with an enormous natural target volume formulated for cosmic-ray neutrinos around the turn of the 1960s, scientific objectives emerging from cosmic-ray theory, and technological solutions developed in international underwater and polar experiments.

The Soviet contribution is especially clear in three areas. The first is the physics and early application of Cherenkov radiation. The second is the formation of underground and underwater high-energy-neutrino detection concepts, including the work of Markov and Zheleznykh and Soviet participation in DUMAND and the Lake Baikal program. The third is the theory of atmospheric and cosmogenic neutrinos, in which work by Zatsepin, Kuzmin, Berezinsky, and their colleagues entered the international scientific tradition.

It would nevertheless be historically inaccurate to describe IceCube as the direct continuation of a Soviet detector or to attribute to Markov in 1960 the proposal for an optical detector in Antarctic ice. The documented early proposal concerned water; ice entered Soviet detector concepts later, predominantly in the radio context; and the optical deep-polar-ice concept was formulated by Halzen and Learned in 1988. This distinction does not diminish the importance of the preceding Soviet work. Rather, it identifies more precisely the role each contribution played in the emergence of modern high-energy neutrino astrophysics.

## Conclusion

The 2026 Nobel Prize in Physics awarded to Francis Halzen marks the maturity of high-energy neutrino astronomy as an observational science. The prehistory of this achievement includes several Soviet research lines that contributed in different ways to the modern field.

The Cherenkov effect, discovered experimentally by P. A. Cherenkov and explained theoretically by I. E. Tamm and I. M. Frank, is the physical basis of optical neutrino telescopes. The work of A. E. Chudakov and N. M. Nesterova in the Pamirs belongs to the early development of Cherenkov methods in cosmic-ray physics and provides historical and methodological context rather than a direct constructional path to IceCube. M. A. Markov formulated in 1960 the principle of large-scale neutrino detection in natural water and, with I. M. Zheleznykh, later developed a broader program of high-energy neutrino detection. G. T. Zatsepin, V. A. Kuzmin, V. S. Berezinsky, and their colleagues established important elements of the theory of atmospheric and cosmogenic neutrinos. Baksan and Lake Baikal became major experimental centers of Soviet and Russian neutrino physics.

The line closest to IceCube in detector logic is the progression from Markov's natural-water concept through DUMAND and Lake Baikal, combined with the later transition to optical detection in polar ice. Soviet radio work in Antarctica formed an important technological and conceptual context, but the direct optical constructional line to IceCube begins with the Halzen–Learned proposal and the AMANDA experiment. The Soviet contribution should therefore be understood not as alternative authorship of IceCube but as a substantial part of the intellectual, theoretical, and experimental history from which modern high-energy neutrino astrophysics emerged.

**Declaration of generative AI and AI-assisted technologies in the writing process**

During the preparation of this manuscript, the author used ChatGPT (OpenAI) and Claude (Anthropic) as AI-assisted tools for translating the article into English, improving language and style, and assisting in the identification and search for relevant bibliographic sources. All AI-assisted outputs were critically reviewed by the author, and the factual content, references, interpretations, and final wording of the manuscript were verified and approved by the author, who takes full responsibility for the published content.